\documentstyle[11pt,IAUS212,twoside,epsf,fleqn]{article}

\def\edcomment#1{\iffalse\marginpar{\raggedright\sl#1\/}\else\relax\fi}
\reversemarginpar

\begin{document}
\vspace*{1cm}
\title{Formation of massive binaries}
 \author{Hans Zinnecker}
\affil{Astrophysikalisches Institut Potsdam, An der Sternwarte 16,\\
14482 Potsdam, Germany}

\begin{abstract}
The formation of massive stars is one of the major
unsolved problems in stellar astrophysics. However,
only few if any of these are found as single stars,
on average massive stars have more than one companion.
Many of them are born in dense stellar clusters and
several clusters have an excess of massive short-period
spectroscopic binaries, with severe implication for
binary-related stellar evolution including mergers,
and also for the origin of massive runaway stars.
The multiplicity of massive stars seems to increase
with increasing primary mass and with increasing
density of young star clusters. These observations suggest
that massive binary and multiple systems originate
mainly from dynamical gravitational interactions
and accretion-induced protostellar collisions
in dense clusters. If true, the binary properties
of massive stars in less dense OB associations should
be less extreme. This prediction should be tested
by future observations.The paper reviews both the latest
observations and theoretical ideas related to the
origin of massive binaries. It
concludes with a speculation on how the binary
properties might change with metallicity (e.g. LMC/SMC).
\end{abstract}

\section{Introduction}

Most massive stars, if not all, are (or were) members of
binary and multiple systems. Ignoring the multiplicity of
massive stars can introduce severe problems when comparing
observations and stellar evolution theory. For example,
the stellar luminosity in the range 20\,--\,60 solar masses
scales with the third power of the mass, and the rate
of production of Lyman continuum photons scales with the
fourth power of the mass. This means that a 50 solar mass
star has 4 times the luminosity of an (unresolved) pair
of two 25 solar mass stars, and 8 times the amount of
Lyman continuum photons per sec. This example, drastic
as it may be, shows the importance of recognising how
the mass of a system is distributed among its components.

The basic properties of stars are imprinted at the time
of stellar formation. Although not the main theme of this
IAU Symposium, any conference on the evolution of massive
stars would be incomplete without considering the very early
formative phases. Of particular interest for predicting the
odyssey of an ensemble of massive stars are the initial
mass function (IMF) and the initial binary star parameters.
While the IMF of massive stars has been a hot topic for a long
time (e.g. Lequeux 1979, Hunter 1995, Scalo 1998, Kroupa 2002),
the binary properties of massive stars have been more of
a side problem, despite several previous spectroscopic studies
(see the brief review in Preibisch, Weigelt, and Zinnecker 2001)

However, an important clue to the origin of high-mass stars
has come from the observation that they occur predominantly in
binary and multiple systems, apparently more
so than low-mass stars (Preibisch, Weigelt, and Zinnecker 2001)
and often with very short orbital periods 
(Mermilliod and Garcia 2001).
Therefore it seems that the understanding of the formation of
high-mass stars is intimately tied to understanding the origin
of their binary nature, be it
via fragmentation of a massive accretion disk 
(Yorke and Sonnhalter 2002),
via Bondi-Hoyle accretion onto a 
lower-mass binary (Maeder and Behrend 2002),
via failed stellar or protostellar collisions (Zinnecker and
Bate 2002) or via dynamical N-body interactions
in dense stellar clusters (Bonnell and Bate 2002).

In this paper, we review the observational situation for
unevolved massive binaries on the Main Sequence in young clusters
and a number of theoretical ideas regarding their formation.
Finally we discuss some implications and predictions,
including a speculation of how the binary properties might
change when we consider low metallicity environments.

\section{Generalities and Motivation}

\subsection{The definition of massive stars and massive binaries}

We believe there is general agreement about the lower mass
limit of a massive star. It is 8 solar masses. Why is
this a good definition? There are at least three reasons:
The first is the fact that stars below 8 solar masses (B3V)
do not emit any Lyman continuum photons. Thus in terms of spectral type
we define massive stars to be earlier than B3V. (A B2.5V star
emits  10$^{44}$\,Lyc photons/sec, an O6V star about 10$^{49}$).
Secondly, for initial stellar masses above 8\,M$_\odot$ there is
no pre-Main Sequence phase (Palla and Stahler 1993), 
i.e. the star ignites H-burning
before it finishes accumulating its final mass by accretion
from the protostellar gas/dust envelope or circumstellar disk.
Thirdly, 8 solar masses is the dividing line between the
initial progenitor mass of white dwarfs and type II supernovae.
Note that the onset of a significant stellar wind mass loss
occurs at around 18 solar masses (B0V), which is the dividing
between O and B stars. Thus only O-stars have stellar winds
on the Main Sequence which can influence their evolution.

The definition of massive binaries is more tricky. The issue
is whether both components must exceed the above 8\,M$_\odot$ limit,
or just the primary of the two. I suggest ``just the primary''.
We will use this latter definition here, especially when it
comes to the search for physical companions to OB-type stars.
Thus a 50\,M$_\odot$\,+\,5\,M$_\odot$ pair or a
10\,M$_\odot$\,+\,1\,M$_\odot$ pair would both be
considered as a massive binary according to our definition.
Of course it will be difficult to detect such pairs as close
spectroscopic binaries but we shall see that it is possible
to resolve them as visual companions, e.g. in the Orion or
M16 young cluster, using speckle interferometry or adaptive
optics.

\subsection{The theme of this contribution}

The underlying theme of this contribution is to study the
massive binary properties as a function of the ``environment'',
in the hope to understand the physics of their formation.
For example, comparison of the properties of massive binaries,
in young clusters vs. OB associations ought to reveal the role
of dynamical gravitational interactions (N-body evolution).
These effects are of great importance in dense star clusters,
while they should play only a minor part in looser aggregates.
Another aspect is the origin of runaway OB stars, such as $\zeta$
Puppis (O4If). Has it been ejected by an asymmetric supernova
explosion or has it been released from a young cluster after
suffering a tight binary-single star encounter? Finally, we
are interested in the initial binary properties as a function
of metallicity, particularly in view of the dynamics of the
protoglobular clusters in the early universe and in the LMC/SMC.
While we will not answer the questions, we want to at least
raise them!

As we are concerned with the formation of massive binaries,
we will concentrate on observations of young unevolved systems
only. The initial or primordial binary properties, as they
are sometimes called, consist of the binary frequency, the
mass ratios, and the component separations.
Measurements of eccentricities are also desirable, as they may
hide clues to formation processes (e.g. tidal capture should
initially result in large eccentricities for close binaries).
One of the key questions is: what is the critical semi-
major axis or critical orbital period for massive
binaries to become interacting binaries later, i.e. for
binary-related Roche lobe mass exchange to occur
when the primary star inflates to become a supergiant?
It depends on the mass ratio of the system, but
for a mass ratio close to unity the critical stellar radius
for Roche lobe overflow is around 0.4 of the semi-major
axis. This implies that massive binaries, with primaries
which can reach radii up to 6\,AU during their post-Main
Sequence evolution, can become semi-detached interacting binaries for
systems with semi-major axes under 15\,AU (i.e. periods
under 10 years). What then is the fraction of all massive
binaries that will turn into interacting ones? My guess
is 25\,--\,50\,\%, considering the orbital period distribution
and mass ratio distribution of bright O-stars found in
the survey of Mason et al. (1998).

\section{Review of observations}

\subsection{Definition of the companion star fraction (CSF)}

Many massive stars are actually members of multiple systems,
e.g. a hierarchical triple or a Trapezium type quadruple.
In order to properly account for these higher order systems,
we introduced the term ``companion star fraction (CSF'')
defined as (B\,+\,2T\,+\,3Q)\,/\,(S\,+\,B\,+\,T\,+\,Q),
see Reipurth and Zinnecker
(1993). Here B is the number of binaries (1 companion),
T the number of triples (2 companions), and Q is the number
of quadruple systems (3 companions). For example:
1 single, 1 binary, 1 triple, and 1 quadruple system in
a cluster of 4 multiple stars like in the Trapezium system
$\theta^1$ Ori (D,C,B,A) in the center of the Orion Nebula cluster
yields CSF\,=\,1.5 (Preibisch et al. 1999).

This means that a massive star in Orion has on average more than
1 companion, which is significantly different from the case
of low-mass pre-Main Sequence stars in Orion,
where we typically find CSF\,=\,0.5 (Prosser et al. 1994,
Padgett et al. 1998).
This discrepancy is the first hint that the formation process of
low-mass and high-mass binary systems may be different.

\subsection{Multiplicity of massive stars in young clusters}

\subsubsection{The Orion Nebula Cluster\\}
 
The multiplicity of the 8 massive stars in the Orion Nebula Cluster
has been studied by Preibisch et al. (1999), using bispectrum
speckle interferometry in the near-infrared K-band. These include
the 4 Trapezium stars $\theta^1$ Ori A, B,C, D, $\theta^2$ Ori A and B 
as well as LP Ori and NU Ori. $\theta^1$ C, the exciting star of
the Orion Nebula has a 33 mas visual companion, previously discovered
by Weigelt et al. (1999). Note that $\theta^1$ C is a peculiar star,
showing strong optical and X-ray variability with a period of
15.4 days (Stahl et al. 1996, Babel \& Montmerle 1997), probably
related to a kG magnetic field misaligned with the rotation axis
(Donati et al. 2002). It also exhibits radial velocity variations,
which have been interpreted as a spectroscopic companion which
could in fact be identical to the visual companion, if the
orbit is sufficiently eccentric (Donati et al. 2002).
$\theta^1$ A and B have long been known to be
spectroscopic binaries (Bossi et al. 1989, Abt et al. 1991), but
in addition they
also have one and two visual companions, respectively. Simon et al.
(1999) even found another infrared companion to $\theta^1$ B, so 
that this system is a Trapezium-type system within the Trapezium.
$\theta^2$ A is a triple system (spectroscopic binary with a 0.2"
visual companion; Abt et al. 1991, Petr et al. 1998);
the same is true for NU Ori. $\theta^1$ D and $\theta^2$ B are apparently
single stars. Adding up the numbers, we find a total of 12 companions,
implying CSF\,=\,12/8\,=\,1.5. 
Interestingly, in all resolved systems
the K-band flux ratios are quite small, i.e less than 1/3. This 
suggests that these visual companions are all significantly less
massive than the primary OB stars.

\subsubsection{The M16 cluster\\}

Duchene et al. (2001) have searched for visual binaries among
the high mass stars in the M16 cluster (NGC 6611), using adaptive
optics. They list some 40 stars of spectral type earlier than B3V 
that are members of the cluster (age 2\,Myr, distance 2\,kpc). 
They find some 10 companions (25\,\%), with separations in the range 
0.5" to 1.5" (1000\,--\,3000\,AU), 
all but two with estimated mass ratios below q\,=\,0.2.
They conclude that these results are consistent with the assumption
that companion masses are randomly drawn from the initial mass
function. The lack of a strong dependence of the properties
of the secondaries on the mass of the primaries lead them to
suggest that the canonical picture of cloud fragmentation can
explain the binary properties, and there is no need to invoke
a different picture for high-mass stars and low-mass stars.
This being said, we caution that the low-mass cluster members provide
a serious source for false binaries (most companions are
physical companions only at the 2 sigma confidence level). 

As for spectroscopic binaries, Bosch, Morrell, and Niemela (1999)
have studied the radial velocities of the 10 earliest type stars
in the NGC 6611 cluster. One short-period (W412, 4 days) and two longer period
SB2s (W175, W197) were found, but three more stars may be variable,
among them the most massive object (W205, O4V).

\subsubsection{NGC 6231 and other clusters\\}

New statistical results about massive spectroscopic binaries
in the young OB cluster were presented by Mermilliod and Garcia (2001)
at the IAU-Symp. 200 (see also Garcia and Mermilliod 2001):

1) The spectroscopic binary frequency among 8 clusters rich in
O-stars (i.e. more than 5 O-star members) varies from extremely high,
of the order of 80\,\%, to rather low, of the order of 15\,\%,
seemingly anti-correlated with stellar density in the clusters.
NGC 6231, a loose cluster,  contains 11 SB out of 14 stars,
while Trumpler 14, the densest cluster in the sample, contains
only 1 SB out of 7 O-stars. An intermediate case is Trumpler 16,
with 7 SBs out of 20 O-star members (35\,\%).

2) Almost all O-stars in the 14 young clusters with few O-stars
(i.e. less than 3 O-star members) are spectroscopic binaries,
often SB2, and even eclipsing. These massive binaries are
usually members of hierarchical triple or quadruple systems,
or of trapezia, and are often located at the cluster center.
$\theta^1$ Ori belongs to this category, although $\theta^1$ C
is not a massive close SB.

3) The orbital periods of the spectroscopic binaries in the
O-star rich clusters are concentrated in the range 4\,--\,5 days,
judging mainly from the NGC 6231 and Trumpler 16 clusters;
while in the O-star poor clusters there is an accumulation
of orbital periods around 3\,$\pm$\,1 day. The orbital
eccentricities are high, in the range 0.2\,--\,0.6, at least
in NGC 6231.

These fascinating and surprising facts challenge our views of
massive binary formation! In the next section, we take up the
challenge, contrasting these clues to some formation models.

\section{Review of theories}

In essence, we can distinguish 4 types of formation theories:

\subsection{Fragmentation of massive disks around massive stars}

Although direct observational evidence for the presence of circumstellar
disks around massive stars is missing (there are indirect hints such as
massive outflows, e.g. Beuther et al. 2002, Henning \& Stecklum 2002),
let us assume that the disks
in question exist, as is known to be the case around low-mass stars.
Support for this scenario comes from the 2D rotating collapse calculations
by Yorke and Sonnhalter (2002); see also Yorke (2002). 
These authors provided simplified numerical
simulations of the accretion of a  massive protostar for a series
of initial cloud cores (30, 60, 120 M$_\odot$) with initially uniform rotation
and no magnetic fields.
Frequency dependent radiative transfer was included. Even with rotation,
radiation pressure on dust is still a limiting factor for the final mass
of the star-disk system. For example, the collapse 
of a 60 M$_\odot$ cloud leads to
a 33 M$_\odot$ star-disk system, with a substantial fraction (about 1/3) still
in the disk. Such a massive disk can be gravitationally unstable,
provided it is cold enough (formally described by the Toomre criterion).
3D simulations would be needed to follow the evolution (fragmentation)
of the disk, but it seems clear that the mass in the disk can only
lead to binary systems with mass ratios less than 1/3, as most of the
mass goes to the central object. Therefore, disk fragmentation cannot
yield nearly equal mass spectroscopic binaries which are so prevalent
among the observed massive binaries. Also, disk fragmentation
produces only wide binaries, of order 10\,--\,100\,AU, comparable to the radial
disk extent (cf. the VLA observations of Shepherd et al. 2001).
This mode may then explain the $\theta^1$ C Ori companion,
but not the SB2 short orbital periods, which correspond to sub-AU
semi-major axes.

\subsection{Bondi-Hoyle accretion onto a low-mass protobinary}

Maeder and Behrend (2002) proposed a model for massive star formation
based on a growing accretion rate of the central star, according to
$\dot{M} \sim M^{1.5}$. 
Realising that the model would have to accommodate
the high incidence and short periods of massive binaries, they
extended their model to include preferential binary formation among
massive stars by Bondi-Hoyle accretion from the protocluster gas
reservoir. Their idea is as follows: The Bondi-Hoyle accretion rate
for an initial binary star (formed by fragmentation or otherwise)
is higher than for a single star of the same mass, particularly if
the Bondi-Hoyle radius of gravitational influence ($b=2 GM/v^2$,
$M$ is the binary mass and $v$ the center-of-mass velocity w.r.t. the gas)
exceeds the semi-major axis of the initial binary system. Then
these systems will ``have more of a chance to move to the top of the
mass scale and they will do so faster than the other single objects.
Since mass is accreted on the protobinary system, the semi-major
axis will decrease and the orbital periods will also decrease''
(Maeder and Behrend 2002).
The problem with this scenario is that the Bondi-Hoyle accretion
rate under normal circumstances (gas density ca. 10$^5$cm$^{-3}$ and
relative velocity ca. 2\,km/s, appropriate for a protocluster cloud)
is far too small to supply the required growth of 10$^{-5}$M$_\odot$\,/yr
for solar mass protostars, unless the relative velocity goes to
zero. This could happen if the orbital motion of the
protobinary partly compensates the relative motion of the gas
w.r.t. the centre-of-mass, and each binary component is
effectively at rest w.r.t. the ambient gas, for part of the
orbit. Clearly the situation is very complicated and detailed
simulations of binary accretion rates are required (cf.
Klessen 2001). Another issue here is whether the Bondi-Hoyle
runaway accretion model can explain the evolution towards
equal mass binary components, perhaps involving a
circumbinary disk (cf. Bate and Bonnell 1997).

\subsection{Encounters and tidal capture of intermediate-mass protostars}

This scenario was originally invented to avoid the radiation pressure
problem of massive star formation (Bonnell, Bate, and Zinnecker 1998).
However, it was soon realised that the collisional build-up of massive
stars could also help to form massive almost-touching eccentric binaries,
if the kinetic energy of a parabolic encounter could be dissipated as
tidal energy (cf. Fabian, Pringle, and Rees 1975) in a gravitationally
focussed grazing fly-by
(Zinnecker and Bate 2002). There are several problems with this
scenario. 

The first one is the extraordinary high stellar density in
the center of a cluster that is needed, so that the collisions
occur at frequent enough intervals to be shorter than the stellar
evolution time of massive stars (few Myr). Stellar number densities
of the order $10^8 stars/pc^3$ are required, which can only be reached,
if a protocluster of intermediate-mass stars undergoes significant
contraction and revirialisation owing to accretion onto the stars
(see Bonnell et al. 1998). A factor of 10 contraction boosts the
stellar number density by a factor of 1000 and shortens the
collision timescale by a factor of 1000. This can happen, but the
question is: does it happen?
No observational support is so far available, although
some hot molecular cores (typical masses ca. 100\,M$_\odot$ and
typical sizes ca. 0.1\,pc) harbour hyper-compact
HII regions (sizes ca. 0.01\,pc), indicative of massive
stars in a very density-peaked gas-rich environment
(Kurtz \& Franco 2002). No-one knows what is going on
inside those very compact radio and mm continuum sources,
as they are so heavily obscured by dust (e.g.,
IRc2-I in Orion OMC1, see Gezari et al. 1998).
It is conceivable that collisional events and 'cannibalism'
are taking place inside, revealed by eruptive phenomena
(e.g., the famous Orion H$_2$ fingers), maser emission,
and thermal infrared flares, as discussed by Bally (2002).

The second problem with this model is the difficulty to
first promote and then avoid stellar mergers. Only in
the last step of this runaway process do we require
the stellar mergers to fail and to form a binary instead.

The third problem is that the formation of massive binaries seems to require
an extreme fine tuning, especially in terms of the impact parameter.
Indeed the cross-section for mergers is always larger than that for
binary formation. One way out is disk formation during the tidal
disruption of the lower-density star in a non head-on collision with
a higher-density star (see the example, shown in Zinnecker and Bate 2002,
of a 3\,M$_\odot$ pre-MS star disrupted by a 10\,M$_\odot$ 
MS-star in a grazing collision).
Disk formation increases the cross-section for dissipative capture
of the next intruder and may also be a general feature of stellar
mergers due to the induced rapid stellar rotation (near break-up
for realistic non-zero impact parameters). We conclude that tidal
capture during near-miss mergers is a viable mechanism to form
massive close binaries, but it requires exceptional circumstances.

Finally, there is the issue whether it is pre-Main Sequence stars
(perhaps with circumstellar disks) or protostellar cores with extended
accretion envelopes that are made to collide (Stahler, Palla, and Ho 2000).
These protostellar envelopes offer a much higher interaction cross-section,
and provide orbital drag for collisional binary formation (Silk 1978).
This variant of the above failed merger picture must be kept in mind,
as it works in less extreme and perhaps more  realistic conditions.

\subsection{N-body interactions in a young cluster}

Bate and Bonnell (2002) recently simulated the gas accretion onto
1000 stars in a gas-rich protocluster using 3D SPH calculations.
Accretion forces the cluster
to contract, leading to a high-density core where stellar mergers
can occur. Small-N groups of massive and intermediate mass stars
form due to the self-gravity of the gas, and binary formation in
these groups is common, occurring through dynamical three-body
capture (different from tidal capture).
To begin with, typically a massive star has a lower-mass wide
companion. With time, an exchange reaction with a more massive
third object occurs, thus forming an almost equal mass but still
wide binary. This wide binary then shrinks (``hardens'') as it
takes up most of the binding energy of the small group being
kicked by the other lower-mass group members most of which tend
to escape from the group. The typical final outcome is a tight
equal mass massive binary, with a lower-mass wide companion
(Bate, Bonnell, and Bromm 2002).
This picture explains many aspects of massive binaries,
including the average companion star fraction ($CSF=1.5$) and
the prevalence of short-period spectroscopic systems.
Also, the timescales for the N-body dynamics are short enough for
the whole process to occur on or before the zero age Main Sequence.
Therefore we favour this scenario as our best working hypothesis.
There is one nagging doubt, however: one would expect the massive
close binary frequency to be correlated with the overall stellar number
density in young clusters, which is not what is observed. In fact,
the contrary (an anti-correlation) is observed. We have no real
explanation at present, how this environmental effect fits into
the N-body picture, unless most of the most massive binaries
in the densest young clusters have managed to merge into a single
object (Zinnecker 1986).

\begin{figure}[h]
\plotone{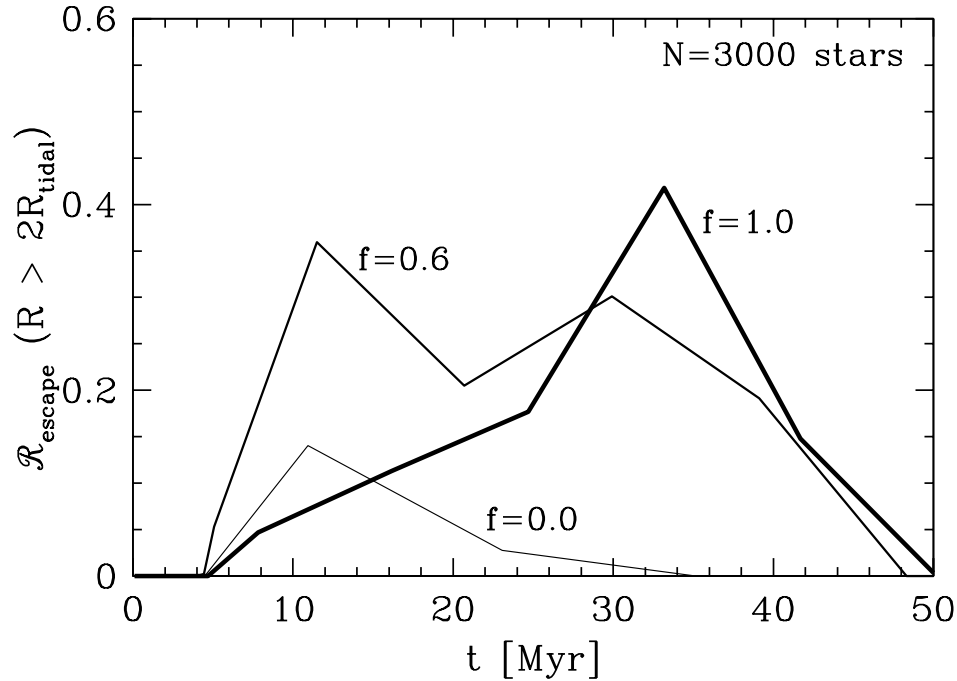}
\caption{Fraction of massive stars that escape a fully
populated cluster of 3000 stars as a function
of time after the first dynamical ejection
from the cluster. The results of numerical
N-body simulations are shown for 3 different
assumptions of the total binary frequency f.
Random pairing from a field star IMF is
assumed to generate the binaries.
The fraction of runaway stars (defined to
be those stars that reach at least twice
the tidal radius of the cluster) is seen to
be 20\,--\,40\,\% between 10\,--\,40\,Myr for realistic
binary frequencies. At later times the
fraction decreases as the massive stars
successively start dying
(courtesy of Kroupa 2000).}
\end{figure}

\section{Implications and Predictions}

Massive binaries may be the main reason for the
existence of many runaway OB stars and field OB stars,
expelled from young clusters. The question whether all
field O-stars (10\,--\,25\,\% of all O-stars) are of some
runaway origin is an extremely crucial one; if not, then
some O-stars must be born in isolation rather than in
clusters, and many conclusion about massive star formation
and indeed massive binary formation must be reconsidered.
For example, the collision scenario
is clearly in trouble if O-stars can form outside clusters.
As it is, however, we know of at least two examples where B0 stars
lie more or less equidistant and in opposite directions
of a young cluster (S255, Zinnecker, McCaughrean, and
Wilking 1993; MonR2, Carpenter 2000).
The prediction (to be tested by future observations) is
that at least one star in each pair of ejected stars must
itself be a binary, to account for the required recoil
energy and momentum; see the simulations of Kroupa (2000)
who also modelled the fraction of massive stars escaping
from a young cluster, thus mimicking an OB field star
population (Fig.\,1).

We end with a speculation on the frequency
and orbital separation of massive binaries in lower
metallicity star forming regions (e.g. LMC/SMC).
We suspect that at lower metallicity the angular momentum
problem of star formation is even more severe than for
molecular clouds with solar composition, implying that
fragmentation must lead to more and/or wider binaries;
perhaps even to faster rotation of the binary components.
The chain of thought goes as follows:
lower metallicity implies fewer heavy elements which
are the main donors of free electrons in quasi-neutral
clouds. Fewer electrons imply a lesser fractional
degree of ionisation. The smaller the degree of
cloud ionisation, the shorter is the timescale of
retaining a magnetic field during cloud contraction
and collapse. Thus the effect of magnetic braking
to help solve the angular momentum problem (e.g.
Mouschovias 1977) is reduced, and the final step
of star formation has to get rid of the extra
angular momentum by forming wider binary systems
in which the remaining excess angular momentum
is stored. If this is true, it will have many
consequences for low-metallicity populations,
but this beyond the scope of this review.

\section*{Questions/Answers:}

\noindent
Peredur Williams:\\
What is the influence of selection effects on the incidence of companions
as a function of type?\\

\noindent
Hans Zinnecker:\\
At present we don't know if there are close binaries with extreme mass
ratios (not detectable as spectroscopic binaries) and we don't know at
all which sort of companions exist in the range of intermediate
separations, too close to be seen as visual binaries in high-spatial
resolution observations and too wide to show up as spectroscopic
binaries. This gap in our knowledge may be bridged to some extent
by future interferometric studies reaching a spatial resolution of the
order of 1\,mas (1\,AU at 1\,kpc distance) and brightness ratios 1\,:\,100.
In summary, I believe that the census of massive stars with lower mass
companions is still quite incomplete.\\

\noindent
Cedric Foellmi:\\
There is an impressive number of scenarios of interacting massive binaries.
We have undertaken in Montreal a systematic search for Wolf-Rayet binaries
in the Magellanic Cloud. We found: ca. 40\,\% binary frequency in the SMC,
15\,--\,25\,\% in the LMC, no evidence for accretion in the secondaries, and
intrinsic hydrogen in the wind of most of the (single and binary) WR stars.
This argues against binary evolution for WR stars where it is theoretically
the most expected (i.e. in a low Z environment).\\

\noindent
Hans Zinnecker:\\
Your new data are very important constraints for stellar evolution and
the close binary frequency of massive stars. However, they do not constrain
the frequency of {\it wide} binaries. In my theoretical speculation, a lower Z
environment tends to favour wider separations of binaries and likely faster
rotation of the components, i.e. in my view a lower Z environment produces
a higher angular momentum population of massive stars. As an aside, I never
claimed that WR stars originate through the binary channel. In fact, this
old concept probably was invented at a time when we had no idea about the
importance of winds from massive stars. (Even at SMC/LMC metallicities
stellar winds can do the trick to form WR stars, without a binary channel).\\

\noindent
Claus Leitherer:\\
I do not see why I would make a large error in the interpretation of data
of unresolved starburst populations if I assumed all stars were single.
Unless stellar evolution changes (which may indeed be the case) very little
should change.\\

\noindent
Hans Zinnecker:\\
It all depends on the frequency of interacting close massive binaries which
can lead to stellar mass exchange and even stellar mergers for the closest
systems. Consequently some rejuvenation of a fraction of the massive stars
can occur, as pointed out by Van Bever and Vanbeveren (1998), implying that
binary evolution can make a starburst look younger than it really is, say
5\,Myr instead of 10\,Myr. A high binary frequency can make a difference here.\\

\noindent
Anatol Cherepashchuk:\\
Ignoring binaries in starbursts will make a big difference for the presence
of compact objects in post-starbursts, and the associated accretion effects
(X-ray binaries).\\

\noindent
Hans Zinnecker:\\
Indeed. Thank you, Mr. chairman.
HZ shows the Chandra X-ray image of the Galactic Centre region
with some 1000 point sources (Wang et al. 2002, Nature 415, 148)

\section*{Acknowledgement}

I would like to thank Karel van der Hucht for his invitation to speak
at IAU-Symp. 212 in Lanzarote and for his patience with the manuscript.
I acknowledge helpful information from Anthony Brown and Pavel Kroupa.
I also owe a great deal of thanks to my massive stars co-investigators
M. Bate, I. Bonnell, Th. Preibisch, G. Weigelt, and also my colleagues
A. Maeder, F. Palla, and H.\,W. Yorke for many stimulating discussions
which helped to shape my ideas on massive star formation.
Finally, I thank the Brazilian participants of the conference to share
with me the experience of the 2002 soccer world-cup final on June 30
(Brazil - Germany 2\,:\,0) in the hotel TV room in Lanzarote.

\end{document}